\documentclass{elsart}

\usepackage{graphicx,amssymb}
\journal{Phys Mech Astron}
\usepackage{graphicx} 
\input{epsf.sty}                        
\input{psfig.sty}

\begin{document}
\begin{frontmatter}
\title{The emission positions of kHz QPOs and Kerr spacetime influence }

\author{ZHANG Chengmin$^{1}$\corauthref{cor}},
\corauth[cor]{Corresponding author.} \ead{zhangcm@bao.ac.cn}
\author{WEI Yingchun$^{1}$}
\author{YIN Hongxing$^{1}$}
\author{ZHAO Yongheng$^{1}$}
\author{LEI YaJuan$^{2}$}
\author{SONG Liming$^{2}$}
\author{ZHANG Fan$^{2}$}
\author{YAN Yan$^{3}$}
\address{1. National Astronomical Observatories, Chinese Academy of Sciences, Beijing 100012, China}
\address{2. Institute of High Energy Physics, Chinese Academy of Sciences, 19B Yuquan Road, Beijing 100049, China}
\address{3.Urumqi Observatory, National Astronomical Observatories, CAS,  Urumqi  830011, China}
\begin{abstract}
\bf Based the Alfven wave oscillation model (AWOM) and
relativistic precession model (RPM) for twin kHz QPOs, we estimate
the emission positions of most detected kHz QPOs to be at
$r=18\pm3 km$ $(R/15km)$ except Cir $X-1$ at $r\sim30\pm5 km
(R/15km)$. For the proposed Keplerian frequency as an upper limit
to kHz QPO, the spin effects in Kerr Spacetime are discussed,
which have about a 5\% (2\%) modification for that of the
Schwarzchild case for the spin frequency of 1000 (400) Hz.The
application to the four typical QPO sources, Cir $X-1$, Sco $X-1$,
SAX J1808.4-3658 and XTE 1807-294, is mentioned.
\end{abstract}

\begin{keyword} kHz QPO, neutron star, low-mass X-ray binaries

\end{keyword}

\end{frontmatter}

\section{\label{sec:1}Introduction}

In thirty more low-mass X-ray binaries (LMXBs), the kiloHertz
quasi-periodic oscillations (kHz QPOs) have been found , where
$2/3$ of them show the twin peak kHz QPOs\cite{01}, upper and
lower frequencies, in the ranges of $\sim100$ Hz - 1300 Hz for the
sources with the different spectrum states, e.g. Atoll and
$Z$\cite{02}. The separations of twin kHz QPOs are not
constant\cite{01,03,04,05,06,07}, which are inconsistent with the
beat model\cite{08,09}. The low frequency QPOs have also been
found, which follow the tight correlations with the kHz QPOs
\cite{01,04}. Some kHz QPO models have been proposed,most of which
are ascribed to the accretion flow\cite{010}, and the Alfven wave
mode oscillation\cite{011,012}. To account for the varied kHz QPO
separation, the relativistic precession model (RPM) is proposed by
Stella and Vietri\cite{013}, which ascribes the upper frequency to
the Keplerian frequency of orbiting material in an accretion disk
and the lower frequency to the periastron precession of the same
matter.

However, for the detected twin kHz QPOs of neutron star (NS) in a
LMXB, their average ratio value is also $3:2$, but varies with the
accretion, which may indicate some distinctions between BHC and
NS\cite{01}. In this short letter, we will investigate the orbital
positions of kHz QPO emissions, based on the Alfven Wave
Oscillation Model (AWOM) \cite{014,015} and RPM\cite{013}. The
Kerr spacetime modification is discussed by considering the spin
influence on the Keplerian frequency.

\section{\label{sec:2}AWOM/RPM for kHz QPOs}

AWOM ascribes an upper frequency to the Keplerian frequency of
orbiting matter at radius r, and a lower frequency to the Alfven
wave oscillation frequency at the same radius, as described in the
following) \cite{014,015},

\begin{equation}
\label{1} \nu_2=\nu_k=1850AX^{\frac{3}{2}}
\end{equation}
with the parameter $X=R/r$ (ratio between star radius R and disk
radius $r$)  and  $A = (m/{R_6^{\ 3}})^{1/2}$ with $R_6 =
R/10^6(cm)$ and $m$ the mass $M$ in the units of solar masses. The
ratio of twin kHz QPO frequencies can be obtained as,
\begin{equation}
\label{2}
{\nu}/{\nu}=(1+(1-x)^{{1/2}})^{\frac{1}{2}}/X^{\frac{4}{5}}
\end{equation}

which only depends on the position parameter $X=R/r$, and has
nothing to do with the other parameters. The twin kHz QPO
separation is obtained as,
\begin{equation}
\label{3}
\nu_2-\nu_1=\nu_2[1-(1-(1-x)^{\frac{1}{2}})^{\frac{1}{2}}]^\ast
X^{\frac{3}{4}}
\end{equation}

\begin{figure}
\begin{center}
\includegraphics[width=12cm]{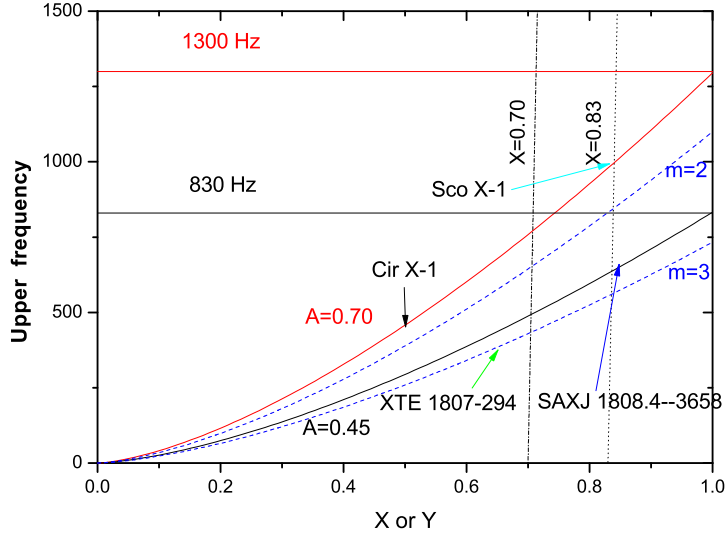}
\end{center}
 \caption
 { Upper kHz QPO frequency vs. the position function $(X=R/r, Y=3Rs/r, Rs = 2GM/C2 )$.
  The upper (down) solid curve represents AWOM with the mass density
parameters $A=0.7$ $(A=0.45)$, where the maximum frequency is
1850A (Hz); The upper (down) dashed curve represents RPM with the
mass parameters $m=2$ $(m=3)$ solar masses, where the maximum
frequency is 2200/m (Hz). } \label{isco-rco}
\end{figure}

\begin{figure}
\begin{center}
\includegraphics[width=12cm]{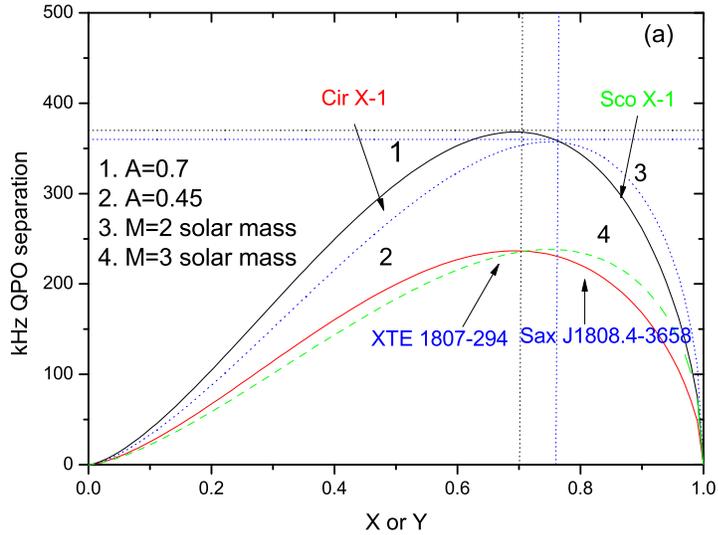}
\end{center}
 \caption
 { Twin kHz QPO separation vs. the position function . Curve 1 and 2
 represent AWOM with  mass density parameters $A=0.7$ and $A=0.45$ respectively.
 Curve 3 (4) represents RPM with  mass parameter $m=2$ $(m=3)$,  respectively.}\label{isco-rco}
\end{figure}

\begin{figure}
\begin{center}
\includegraphics[width=12cm]{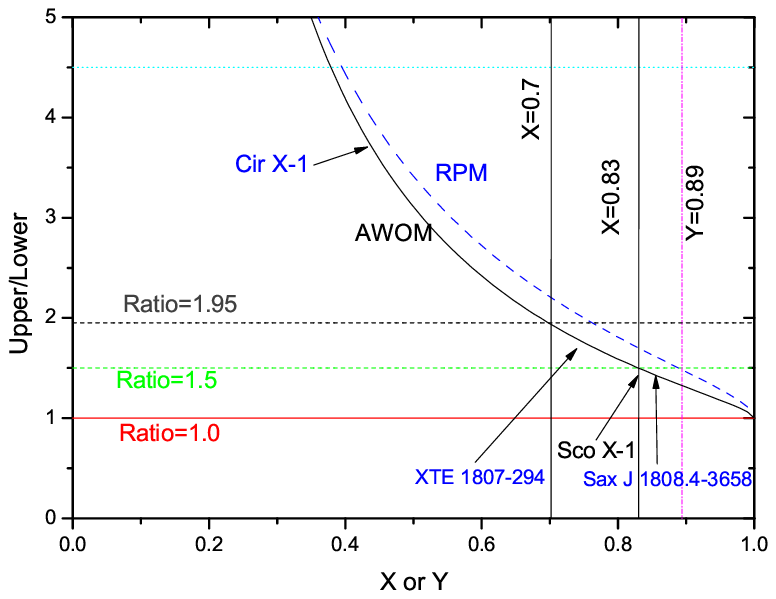}
\end{center}
 \caption
 { Twin kHz QPO ratio vs. the position function (Same meaning as shown in FIG.1).
 The ratio 1.5 (1) is the averaged (minimum limit) value of the detected twin
 kHz QPOs.}\label{isco-rco}
\end{figure}

In FIG.1, the upper kHz QPO frequency is plotted against the
position parameter $X=R/r$ ($Y=3Rs/r$,  $Rs$ is the Schwarzschild
radius) for AWOM (RPM). For the detected twin kHz QPOs, the mass
density parameter A is found to be about 0.7 (e.g. Sco $X-1$)
\cite{014,015}.  In most cases (except Cir $X-1$), the position
parameter $X =R/r$ is lies in the range from 0.7 to 0.92, or
radius from $r=1.1R$ to $r=1.4R$. This implies that the emission
positions of most kHz QPOs are close to  the surface of the NS
$X=1$ for AWOM (for RPM the emission positions are close to
$3Rs$), which means that the maximum kHz QPO frequency occurs at
the surface (ISCO of star $r=R$ for AWOM (or $r=3Rs$ for RPM).  In
FIG.2, the twin kHz QPO separation vs. position parameter is
plotted, where the maximum separation 375 (200) Hz is achieved for
$A=0.7$ (0.45) at $X=0.7$ for AWOM. The kHz QPO data of two
accretion powered millisecond X-ray pulsars (AMXPs), Sax J
1808.4-3658 and XTE 1807-294, approximately hint at the condition
of $A=0.45$, which presents relatively low kHz QPO separations.
For RPM, the maximum kHz QPO separations are 360 Hz (210 Hz) for
the different choices of mass parameter $m=2 (3)$ solar masses,
which occurs at $Y=0.76$. For the two AMXPs, Sax J 1808.4-3658 and
XTE 1807-294, RPM has to assume their star masses are close to the
NS mass upper limit, 3 solar masses, if consistent theoretical
curves with the detected data can be fitted.  FIG.3 is the diagram
of twin kHz QPO ratios vs. position parameter. It can be noticed
that the averaged ratio 1.5 of the detected kHz QPOs corresponds
to the position $X=0.83$ for AWOM ($Y=0.89$ for RPM). The ratios
of all sources but Cir $X-1$ lie in the regimes between $ratio=1$
and $ratio=2$. The kHz QPO data of Cir $X-1$ implies that its kHz
QPO emitting positions are far away from the star, i.e. $0.4 < X <
0.6$ or $2.5R
> r > 1.6R$, centered at about $2R$.

\section{\label{sec:3}Kerr spacetime effect on the kHz QPO}

If the influence of Kerr spacetime on the Keplerian frequency is
taken into account, then the orbital frequency of a spinning point
mass $M$ with angular momentum $J$ is expressed as below\cite{01}
\begin{equation}
\label{4} \nu_2=\nu\nu_k\xi;\quad \nu_k=(GM/4\Pi
r^3)^{\frac{1}{2}}
\end{equation}
with the Kerr modification parameter
\begin{equation}
\label{5} \xi=1+jR_g^{\ \frac{2}{3}};\quad R_g=R_s/2
\end{equation}

\begin{equation}
\label{6} j=Jc/GM^2; \quad J=2\pi I\nu_s
\end{equation}
where $I$ is the moment of inertia, with the maximum value for the
homogeneous sphere $I = (2/5)MR^2$. In the Schwarzschild geometry,
$j = 0$, Eq.\ref{4} recovers the conventional Keplerian frequency;
$0 < j < 1$ represents a prograde orbit. To put the NS mass
$(m=M/M_\odot$, radius and spin fre-quency parameters, we have the
following simplified expressions,

\begin{equation}
\label{7} j=4\Pi\nu_2R^2/R_gC=(0.22/m)R_6^{\ 2}(\nu_s/400HZ)
\end{equation}

\begin{equation}
\label{8} \xi=1+(0.0013m)R_6(\nu_s/400hz)
\end{equation}

If we set the conventional values $M=1.4 M_\odot$, $R=15 km$ and
$s = 400Hz$, then the Kerr modification parameter has about a 2\%
contribution to the Keplerian frequency, which cannot have too
much influence on the kHz QPO model based on the Keplerian
frequency. For the maximum spin fre-quency $1122 Hz$, the Kerr
modification contributes about 5\% to the Schwarzschild spacetime,
so this influence should be considered when we estimate the NS
parameters.

\section{\label{sec:4}Discussions and results}
The kHz QPO emission positions are analyzed by the models (AWOM
and RPM), which shows that most kHz QPOs (e.g. Sco $X-1$) come
from the regimes of several kilometers away from the stellar
surface. This may correspond to the condition of a spinning up NS,
since the detected NS spin frequencies are averaging $400
Hz$\cite{016}, less than the upper frequencies. In RPM, the star
mass can be derived by the detected twin kHz QPOs, then it usually
gives a value of 2 solar masses, higher than the typical NS mass
of 1.4 solar masses. One reason for RPM's prediction of high NS
mass may be originating from its assumption of the vacuum
circumstance around the star in introducing the perihelion
precession term\cite{013}, but the accretion disk does not satisfy
this clean condition. A value of about 3 solar masses for SAX
J1808.4-3658 \cite{017} (e.g. XTE 1807-294) is obtained, which
seems to suggest that RPM should be modified. AWOM cannot predict
a stellar mass by QPO but rather an averaged mass density $(A\sim
M^{1/2}/R^{3/2)}$, by which one can evaluate the equation of state
(EOS) of the star. For the presently known kHz QPO frequencies,
AWOM cannot give the prediction of quark matter\cite{018} inside
the star unless the QPO frequency over 1500 Hz is detected. In
addition, the Kerr spacetime influence is investigated, and a 5\%
modification factor in Keplerian frequency exists for a high spin
frequency of 1000 Hz, which will increase the estimation of the
mass density parameter. Though the spectral properties of Cir
$X-1$ are typical of those of Z sources\cite{019,020}, its
detected  11 pairs of kHz QPOs are generally low frequencies, 230
Hz to 500 Hz for the upper QPO and 56 Hz to 225 Hz for the lower
QPO, increasing with accretion rate, which is contrary to those of
the other LMXBs. The peak separation lies at 175-340 Hz, similar
to those of other LMXBs \cite{06}. Since the kHz QPO emitting
positions of Cir $X-1$ are estimated to be beyond the orbit of 25
kilometers, we guess that its rotating frequency is low, e.g. a
hundred Hz.

\ack{ This work is supported by National Basic Research Program of
China-973 Program 2009CB824800; National Natural Science
Foundation of China (10773017).}

\end{document}